\documentclass[aps,preprint]{revtex4}%
\usepackage{amsfonts}
\usepackage{amsmath}
\usepackage{amssymb}
\usepackage{graphicx}%
\providecommand{\U}[1]{\protect\rule{.1in}{.1in}}

\begin{document}
\preprint{ }
\title{Quantum histories without contrary inferences}
\author{Marcelo Losada}
\affiliation{Instituto de F\'{\i}sica Rosario, Rosario, Argentina}
\author{Roberto Laura}
\affiliation{Instituto de F\'{\i}sica Rosario and Facultad de Ciencias Exactas,
Ingenier\'{\i}a y Agrimensura, Rosario, Argentina}

\begin{abstract}
In the consistent histories formulation of quantum theory it was shown that it
is possible to retrodict contrary properties. We show that this problem do not
appear in our formalism of generalized contexts for quantum histories.

\end{abstract}
\date{January 2014}
\maketitle

\section{Introduction.}

In the consistent histories formulation of quantum theory \cite{Gri}
\cite{Omn} \cite{Gell}, the probabilistic predictions and retrodictions depend
on the choice of a consistent set. It was shown that this freedom allows the
formalism to retrodict two contrary properties \cite{Kent}. This is not a
problem for the defenders of the theory, because each retrodiction is obtained
in a different consistent sets of histories, i.e. in different descriptions of
the physical system not to be considered simultaneously \cite{Har}
\cite{GriHar}. However, this fact is considered by some authors as a serious
failure of the theory of consistent histories \cite{Kent} \cite{Sud}
\cite{Kent2}.

We are going to analyze this problem with our formalism of generalized
contexts \cite{con leo} \cite{con marcelo}, developed to deal with expressions
involving properties at different times. The formalism is an alternative to
the theory of consistent histories, which has proved to be useful for the time
dependent description of the logic of quantum measurements \cite{medicion},
the decay processes \cite{decaimiento} and the double slit experiment with and
without measurement instruments \cite{con marcelo}. More recently \cite{aop}
we have discussed the relation of our formalism with the theory of consistent histories.

In section II we show that there is no possibility for contrary inferences in
ordinary quantum mechanics. In section III we discuss the retrodiction of
contrary properties in the theory of consistent histories. In section IV we
show that there are no retrodiction of contrary properties in our formalism of
generalized contexts. The main conclusions are given in section V.

\section{Contrary properties in an ordinary quantum context.}

In quantum mechanics, a property $p$ is represented by a projector $\Pi_{p}$
in the Hilbert space $\mathcal{H}$, or alternatively by the corresponding
Hilbert subspace $V_{p}=\Pi_{p}\mathcal{H}$. By definition \cite{Kent}, two
quantum properties $p$ and $q$ are said to be \textit{contrary} if they
satisfy the order relation $p\leq\overline{q}$, which can also be expressed in
terms of the inclusion of the corresponding Hilbert subspaces in the form
\begin{equation}
\Pi_{p}\mathcal{H}\subseteq(I-\Pi_{q})\mathcal{H}. \label{contrary}%
\end{equation}

The inclusion of subspaces is equivalent to the following relation between the
corresponding projectors (see \cite{Mit}, section 1.3)%
\[
\Pi_{p}(I-\Pi_{q})=(I-\Pi_{q})\Pi_{p}=\Pi_{p},
\]
from which we easily deduce that $\Pi_{p}\Pi_{q}=\Pi_{q}\Pi_{p}=0$, that means
the projectors $\Pi_{p}$ and $\Pi_{q}$ are orthogonal.

As they also commute, $p$ and $q$ are compatible properties. The projectors
$\Pi_{p}$, $\Pi_{q}$ and $\Pi_{\overline{p\vee q}}=I-\Pi_{p}-\Pi_{q}$ form a
projective decomposition of the Hilbert space, i.e. they are orthogonal and
their sum is the identity operator. Therefore, the properties $p$, $q$ and
$\overline{p\vee q}$ can be considered the atomic properties generating a
\textit{context} of quantum properties with well defined probabilities
\cite{con marcelo}.

For any state of the system represented by a state operator $\rho$, the
probability of any property $p^{\prime}$ in the context is obtained with the
Born rule, i.e. $\Pr_{\rho}(p^{\prime})=$Tr$(\rho\Pi_{p^{\prime}})$. For the
atomic properties $p$, $q$ and $\overline{p\vee q}$ we obtain%
\begin{equation}
\Pr_{\rho}(p)+\Pr_{\rho}(q)+\Pr_{\rho}(\overline{p\vee q})=1 \label{suma}%
\end{equation}
From this equation we easily deduce that if $\Pr_{\rho}(p)=1$ then $\Pr_{\rho
}(q)=0$ and if $\Pr_{\rho}(q)=1$, $\Pr_{\rho}(p)=0$.

We conclude that in ordinary quantum mechanics it is impossible for any state
$\rho$ that two contrary properties $p$ and $q$ have probability equal to one.
These results are the stochastic version of contrary proposition in ordinary
logic. They can be interpreted by saying that whenever the property $p$ ($q$)
is true, the property $q$ ($p$) is false. By the way, this result also justify
to have given the name \textit{contrary} to quantum properties $p$ and $q$
satisfying equation (\ref{contrary}).

More generally, it is easy to see that if $p$ and $q$ are contrary properties,
it is not possible to have a state $\rho$ and another property $r$ for which%
\begin{equation}
\Pr_{\rho}(p|r)=1,\qquad\Pr_{\rho}(q|r)=1. \label{condicional}%
\end{equation}

Taking into account that $p$, $q$ and $r$ should be represented by commuting
projectors, so that the conditional probabilities be well defined, we would
have%
\[
\Pr_{\rho}(p|r)=\frac{\text{Tr}(\rho\Pi_{p}\Pi_{r})}{\text{Tr}(\rho\Pi_{r}%
)}=\text{Tr}(\rho^{\ast}\Pi_{p})=\Pr_{\rho^{\ast}}(p),\qquad\Pr_{\rho
}(q|r)=\text{Tr}(\rho^{\ast}\Pi_{q})=\Pr_{\rho^{\ast}}(q),
\]
where $\rho^{\ast}\equiv\frac{\Pi_{r}\rho\Pi_{r}}{\text{Tr}(\Pi_{r}\rho\Pi
_{r})}$. Taking into account equation (\ref{suma}) with $\rho=\rho^{\ast}$ we
conclude that there are no state $\rho$ and property $r$ for which equations
(\ref{condicional}) can be both valid.

\section{Contrary properties in the theory of consistent histories.}

In the theory of consistent histories $n$ different contexts of properties at
each time $t_{j}$ $(j=1,...,n)$, satisfying a state dependent consistency
condition, can be used to define a \textit{family of consistent histories},
i.e. a set of $n$ times sequences of properties with well defined
probabilities \cite{Gri} \cite{Omn} \cite{Gell}. According to the theory, each
possible family of consistent histories is an equally valid description of the
quantum system. In general it is not possible to include two different
families in a single larger one. Different families of this kind are
complementary descriptions of the system, which the theory excludes to be
considered simultaneously.

A discussion on the logical aspects of the theory was opened by Adrian Kent
\cite{Kent}, who first pointed out that it is possible the retrodiction of
contrary properties in different families of consistent histories, i.e.%
\begin{equation}
\Pr_{\rho_{t_{0}}}(p,t_{1}|r,t_{2})=1,\qquad\Pr_{\rho_{t_{0}}}(q,t_{1}%
|r,t_{2})=1, \label{pro}%
\end{equation}
where $\rho_{t_{0}}$ is the state of the system at time $t_{0}$, $p$ and $q$
are contrary properties at time $t_{1}>t_{0}$ and $r$ is a property at time
$t_{2}>t_{1}$ (see references \cite{Kent} and \cite{Har} for explicit
expressions of $p$, $q$, $r$ and $\rho_{t_{0}}$).

The first equation above is valid for the consistent family that includes $p$
and $\overline{p}$ at time $t_{1}$ together with $r$ and $\overline{r}$ at
time $t_{2}$. It gives the retrodiction of property $p$ at time $t_{1}$
conditional to property $r$ at time $t_{2}$. The second equation is valid for
the consistent family including $q$ and $\overline{q}$ at time $t_{1}$
together with $r$ and $\overline{r}$ at time $t_{2}$, and it gives the
retrodiction of property $q$ at time $t_{1}$ conditional to property $r$ at
time $t_{2}$.

From the point of view of the theory of consistent histories equations
(\ref{pro}) cannot be interpreted as the retrodiction of two contrary
properties, because they are valid in two different and complementary
descriptions, which cannot be included in a single consistent family
\cite{GriHar} \cite{Har}. However, some authors have considered the results
given in equations (\ref{pro}) as a serious objection for the internal
consistency of the theory of consistent histories \cite{Kent} \cite{Kent2}
\cite{Sud}.

\section{Contrary properties in the formalism of generalized contexts.}

In this section contrary quantum properties will be considered from the point
of view of our formalism of generalized contexts. We start with a brief
description of the formalism, which was presented in full details in our
previous papers \cite{con leo} \cite{con marcelo}.

Quantum mechanics do not give a meaning to the joint probability distribution
of observables whose operators do not commute. It can only deal with a set of
properties belonging to a context.

A context of properties $\mathcal{C}_{i}$ at time $t_{i}$ is obtained starting
from a set of atomic properties $p_{i}^{k_{i}}$ $(k_{i}\in\sigma_{i})$
represented by projectors $\Pi_{i}^{k_{i}}$ corresponding to a projective
decomposition of the Hilbert space $\mathcal{H}$, i.e. verifying
\begin{equation}%
{\textstyle\sum\nolimits_{k_{i}\in\sigma_{i}}}
\Pi_{i}^{k_{i}}=I,\qquad\Pi_{i}^{k_{i}}\Pi_{i}^{k_{i}^{\prime}}=\delta
_{k_{i}k_{i}^{\prime}}\Pi_{i}^{k_{i}}. \label{A}%
\end{equation}
Any property $p$ of the context $\mathcal{C}_{i}$ is represented by a sum of
the projectors of the projective decomposition,
\begin{equation}
\Pi_{p}=%
{\textstyle\sum\nolimits_{k_{i}\in\sigma_{p}}}
\Pi_{i}^{k_{i}},\qquad\sigma_{p}\subset\sigma_{i}. \label{B}%
\end{equation}

The context $\mathcal{C}_{i}$ is an orthocomplemented distributive lattice,
with the complement $\overline{p}$ of a property $p$ defined by $\Pi
_{\overline{p}}\equiv I-\Pi_{p}$ and the order relation $p\leq p^{\prime}$
defined by $\Pi_{p}\mathcal{H\subseteq}\Pi_{p^{\prime}}\mathcal{H}$.

A well defined probability (i.e. additive, non negative and normalized) is
defined by the Born rule $\Pr_{t_{i}}(p)\equiv$Tr$(\rho_{t_{i}}\Pi_{p})$ on
the context $\mathcal{C}_{i}$. In Heisenberg representation, the probability
of a property $p$ at time $t_{i}$ can be written in terms of the state at a
reference time $t_{0}$, i.e.
\begin{equation}
\Pr_{t_{i}}(p)=Tr(\rho_{t_{0}}\Pi_{p,0}),\qquad\Pi_{p,0}\equiv U(t_{0}%
,t_{i})\Pi_{p}U(t_{i},t_{0}),\qquad U(t_{i},t_{0})=e^{-\frac{i}{\hslash
}H(t_{i}-t_{0})}.\label{C}%
\end{equation}

Taking into account equations (\ref{B}) and (\ref{C}), the Heisenberg
representation of the property $p$ of the context $\mathcal{C}_{i}$ at time
$t_{i}$ is given by%
\begin{equation}
\Pi_{p,0}=%
{\textstyle\sum\nolimits_{k_{i}\in\sigma_{p}}}
\Pi_{i,0}^{k_{i}},\label{D}%
\end{equation}
where the projectors $\Pi_{i,0}^{k_{i}}=U(t_{0},t_{i})\Pi_{i}^{k_{i}}%
U(t_{i},t_{0})$ represent the time translation of the atomic properties
$p_{i}^{k_{i}}$ from time $t_{i}$ to the time $t_{0}$. The projectors
$\Pi_{i,0}^{k_{i}}$ also satisfy equations (\ref{A}).

The Heisenberg representation of the context $\mathcal{C}_{i}$ at time $t_{i}$
suggest a generalization of quantum mechanics for including the joint
probability of properties belonging to different contexts $\mathcal{C}_{1}%
$,...,$\mathcal{C}_{i}$,...,$\mathcal{C}_{n}$ corresponding to $n$ different
times $t_{1}<...<t_{i}<...<t_{n}$.

By extending what is a common assumption in ordinary quantum mechanics, we
proposed to give a meaning to the joint probability of properties at different
times if they correspond to commuting projectors in Heisenberg representation.
This will be the case if the atomic properties generating each of the $n$
contexts are represented by projectors satisfying%
\[
\lbrack\Pi_{i,0}^{k_{i}},\Pi_{j,0}^{k_{j}}]=0,\qquad i,j=1,...,n,\qquad
k_{i}\in\sigma_{i},\qquad k_{j}\in\sigma_{j}.
\]

If these projectors commute the projectors $\Pi_{0}^{\mathbf{k}}\equiv
\Pi_{1,0}^{k_{1}}...\Pi_{i,0}^{k_{i}}...\Pi_{n,0}^{k_{n}}$, with
$\mathbf{k}=(k_{1},...,k_{n})$ and $k_{i}\in\sigma_{i}$, form a projective
decomposition of the Hilbert space $\mathcal{H}$, as they satisfy%
\[%
{\textstyle\sum\limits_{\mathbf{k}}}
\Pi_{0}^{\mathbf{k}}=I,\qquad\Pi_{0}^{\mathbf{k}}\Pi_{0}^{\mathbf{k}^{\prime}%
}=\delta_{\mathbf{kk}^{\prime}}\Pi_{0}^{\mathbf{k}},\qquad\mathbf{k}%
,\mathbf{k}^{\prime}\in\sigma_{1}\times...\times\sigma_{n}%
\]

In our formalism we postulate that an expression of the form \textquotedblleft%
\textit{property} $p_{1}^{k_{1}}$ \textit{at time} $t_{1}$ \textit{and} ~
\textit{...} ~ \textit{and} $p_{n}^{k_{n}}$ \textit{at time} $t_{n}%
$\textquotedblright\ is an atomic generalized property $\mathbf{p}%
^{\mathbf{k}}$ with the Heisenberg representation given by the projector
$\Pi_{0}^{\mathbf{k}}$. A \textit{generalized context} is defined by all the
generalized properties $\mathbf{p}$ having a Heisenberg representation given
by an arbitrary sum of the projectors $\Pi_{0}^{\mathbf{k}}$, i.e.%
\[
\Pi_{\mathbf{p}}=%
{\textstyle\sum\limits_{\mathbf{k}\in\sigma_{\mathbf{p}}}}
\Pi_{0}^{\mathbf{k}},
\]
where $\sigma_{\mathbf{p}}$ is a subset of $\sigma_{1}\times...\times
\sigma_{n}$. The generalized context is an orthocomplemented distributive
lattice, with the complement $\overline{\mathbf{p}}$ of $\mathbf{p}$ defined
by $\Pi_{\overline{\mathbf{p}}}=I-\Pi_{\mathbf{p}}$, and the order relation
$\mathbf{p}\leq\mathbf{p}^{\prime}$ defined by the inclusion of the
corresponding Hilbert subspaces ($\Pi_{\mathbf{p}}\mathcal{H\subseteq}%
\Pi_{\mathbf{p}^{\prime}}\mathcal{H}$).

An extension of the Born rule provides a definition of an additive, non
negative and normalized probability on the generalized context, given by%
\begin{equation}
\Pr(\mathbf{p})\equiv\text{Tr}(\rho_{t_{0}}\Pi_{\mathbf{p}}). \label{E}%
\end{equation}

We are now going to analyze the retrodiction of contrary properties in the
formalism of generalized contexts. We consider a state $\rho_{t_{0}}$ at time
$t_{0}$, two contrary properties $p$ and $q$ at time $t_{1}>t_{0}$ and another
property $r$ at time $t_{2}>t_{1}$, and we search for the possibility to
obtain for both conditional probabilities the results $\Pr_{\rho_{t_{0}}%
}(p,t_{1}|r,t_{2})=1$ and $\Pr_{\rho_{t_{0}}}(q,t_{1}|r,t_{2})=1$.

The projectors $\Pi_{p}$ and $\Pi_{p,0}=U(t_{0},t_{1})\Pi_{p}U(t_{1},t_{0})$
are respectively Schr\"{o}dinger and Heisenberg representations of the
property $p$ at time $t_{1}$. Analogously, $\Pi_{q}$ and $\Pi_{q,0}%
=U(t_{0},t_{1})\Pi_{q}U(t_{1},t_{0})$ are representations of the property $q$
at time $t_{1}$. Moreover, $\Pi_{r}$ and $\Pi_{r,0}=U(t_{0},t_{2})\Pi
_{r}U(t_{2},t_{0})$ are representations of the property $r$ at time $t_{2}$.

The conditional probabilities are meaningful in our formalism if the following
compatibility conditions are satisfied%
\begin{equation}
\lbrack\Pi_{p,0},\Pi_{r,0}]=0,\qquad\lbrack\Pi_{q,0},\Pi_{r,0}]=0,\label{co1}%
\end{equation}
while the contrary properties $p$ and $q$ are represented by orthogonal
projectors, and therefore%
\begin{equation}
\lbrack\Pi_{p,0},\Pi_{q,0}]=0.\label{co2}%
\end{equation}

The commutation relations given in equations (\ref{co1}) and (\ref{co2}) are
the compatibility conditions required to consider a two times generalized
context including the contrary properties $p$ and $q$ at time $t_{1}$ and
property $r$ at time $t_{2}$, in which both conditional probabilities
$\Pr_{\rho_{t_{0}}}(p,t_{1}|r,t_{2})$ and $\Pr_{\rho_{t_{0}}}(q,t_{1}%
|r,t_{2})$ are meaningful.

In our formalism, the required retrodictions would have the explicit forms%
\[
\Pr_{\rho_{t_{0}}}(p,t_{1}|r,t_{2})=\frac{\text{Tr}(\rho_{t_{0}}\Pi_{p,0}%
\Pi_{r,0})}{\text{Tr}(\rho_{t_{0}}\Pi_{r,0})}=1,\qquad\Pr_{\rho_{t_{0}}%
}(q,t_{1}|r,t_{2})=\frac{\text{Tr}(\rho_{t_{0}}\Pi_{q,0}\Pi_{r,0})}%
{\text{Tr}(\rho_{t_{0}}\Pi_{r,0})}=1.
\]

Taking into account the commutation relations given in equations (\ref{co1}),
the previous equations are equivalent to%
\begin{equation}
\text{Tr}(\rho_{t_{0}}^{\ast}\Pi_{p,0})=1,\qquad\text{Tr}(\rho_{t_{0}}^{\ast
}\Pi_{q,0})=1,\qquad\rho_{t_{0}}^{\ast}\equiv\frac{\Pi_{r,0}\rho_{t_{0}}%
\Pi_{r,0}}{\text{Tr}(\Pi_{r,0}\rho_{t_{0}}\Pi_{r,0})}.\label{both}%
\end{equation}

As $\Pi_{p,0}$ and $\Pi_{q,0}$ represent contrary properties at the same time
$t_{0}$, we can follow the arguments given at the end of section II to show
that there is no $\rho_{t_{0}}^{\ast}$ for which both equations given in
equations (\ref{both}) can be valid. Therefore we conclude that the problem of
retrodiction of contrary properties do not arise in our formalism of
generalized contexts for quantum histories

\section{Conclusions.}

In ordinary quantum mechanics, contrary properties are represented by
orthogonal subspaces of the Hilbert space associated with the physical system.
In section II, we proved that given two contrary properties $p$ and $q$, there
is no state $\rho$ and property $r$ for which the probability of $p$
conditional to $r$ and the probability of $q$ conditional to $r$ can be both
equal to one. Therefore, there is no possibility of contrary inferences in
ordinary quantum mechanics. This result corresponds to a state and properties
considered at a single time.

As we discussed in section III, this is not the case for the theory of
consistent histories, where a state at time $t_{0}$, two contrary properties
$p$ and $q$ at time $t_{1}>t_{0}$ and another property $r$ at time
$t_{2}>t_{1}$ can be found in such a way that the probability of $p$
conditional to $r$ and the probability of $q$ conditional to $r$ are both
equal to one. Although these conditional probabilities are defined in
different sets of consistent histories \cite{Har} \cite{GriHar}, some authors
have considered this fact as a serious problem for the logical consistency of
the theory \cite{Kent} \cite{Sud} \cite{Kent2}.

The main purpose of this paper was to analyze the problem of contrary
inferences in the framework of our formalism of generalized contexts. In this
formalism, as it was explained in section IV, ordinary contexts of properties
at different times can be used to obtain a valid set of quantum histories if
they satisfy a compatibility condition. This condition is given by the
commutation of the projectors corresponding to the time translation of the
properties to a single common time. These compatibility conditions are state
independent, an important difference with respect to the state dependent
consistency conditions of the theory of consistent histories. Each quantum
history has a Heisenberg representation given by a projection operator, and
each valid set of quantum histories is generated by a projective decomposition
of the Hilbert space. As a consequence, a generalized context of quantum
histories has the logical structure of a distributive orthocomplemented
lattice of subspaces of the Hilbert space, i.e. the same logical structure of
the quantum properties of an ordinary context. It is because of this logical
structure that in our formalism there is no place for the retrodiction of
contrary properties.

Recently we have annalyzed the relations of our formalism with the theory of
consistent histories \cite{aop}. Our formalism was also successful in
describing the time dependent logic of quantum measurements \cite{medicion},
the quantum decay process \cite{decaimiento} and the double slit experiment
with and without measurement instruments \cite{con marcelo}. The results of
this paper encourages us to continue our future research considering more
applications of the formalism of generalized contexts.

\end{document}